\documentstyle[prb,aps,epsf,multicol]{revtex}

\begin{document}

\title{ Ballistic spin-polarized transport and Rashba spin
         precession in semiconductor nanowires}

\author{ Francisco Mireles and George Kirczenow}

\address { Department of Physics, Simon Fraser University, Burnaby,
            BC VSA 1S6, Canada.}

\date{May 3, 2001}


\maketitle

\begin{abstract}
We present numerical calculations of the ballistic spin-transport
properties of quasi-one-dimensional wires  in the presence of the
spin-orbit (Rashba) interaction. A tight-binding analog of the
Rashba Hamiltonian which  models the Rashba effect is used. By
varying  the robustness of the Rashba coupling and the width of the
wire,  {\it
weak} and {\it strong} coupling regimes  are identified. Perfect electron
spin-modulation is found for the former  regime, regardless of the
incident Fermi energy and mode number. In the latter however, the
spin-conductance  has  a strong energy dependence  due to a
nontrivial subband intermixing induced by the strong  Rashba
coupling. This would imply a strong suppression of the
spin-modulation at higher temperatures and source-drain voltages.
The results may be of relevance for the implementation of
quasi-one-dimensional spin transistor devices.
\end{abstract}

\pacs
 

\begin{multicols}{2}
\section{INTRODUCTION}

The electron spin precession phenomena at zero magnetic field induced
by a variable spin-orbit interaction in quasi-two-dimensional
electron gas (2DEG) systems, was first elucidated by Datta and
Das\cite{Datta-Das} as the basic principle for the  realization of a
novel electronic device, the spin-transistor. The  underlying  idea
is to drive a modulated spin-polarized current (spin-inject and
detect)  entirely  electrically, combining just ferromagnetic metals
and semiconductor  materials.  For this, the  spin-precession is
controlled via the spin-orbit coupling (Rashba-coupling\cite
{Rashba,Bychkov-Rashba}) associated with the interfacial electric
fields present  in the quantum well  that contains the 2DEG
channel. The tuning of the Rashba coupling  by an external
gate voltage, was recently demonstrated in different
semiconductors by Nitta {\it et  al.}\cite{Nitta} and
others\cite{Heida,Engels,Hu}, and  more recently  by
Grundler\cite{Grundler}  applying a back gate voltage while the
carrier density was kept constant. It has been also achieved in
a {\it  p-}type InAs semiconductor by Matsuyama {\it et
al.}\cite{Matsuyama}

Although spin injection has already been  reported from
ferromagnetic metals to InAs based semiconductors,
\cite{Hammar,Gardelis} the spin polarization signatures reported are
about  1\% or less, making  the results very controversial. Such
low efficiency can be also attributed to   extraneous effects, such
as the local Hall effect,\cite{Filip,Tang} and to resistance mismatch
at the interfaces between the ferromagnetic metal and the
semiconductor.\cite{Schmidt} Very recently, one of us has proposed
that the latter problem can be circumvented by growing atomically
ordered and appropriately  oriented interfaces of ferromagnetic
metals and suitable  semiconductors, that act as perfect
spin-filters and make injection of up to $100\%$ spin-polarized
electrons into semiconductors possible in principle.
\cite{G-Kirczenow} If this prediction is confirmed experimentally, a
major obstacle to the spin injection experiments will  be overcome.

On the other hand, another crucial prerequisite to having an overall
strong spin current modulation, is to restrict the angular
distribution of electrons in the 2DEG by imposing a strong enough,
transverse confining potential.\cite{Datta-Das} This was the original
proposal of Datta and Das. It was argued that if $W$ is the width of
the  transverse confining potential well, the condition
$W\ll\hbar^2/\alpha m^*$, should be satisfied for the intersubband
mixing to be negligible, which would ensure a perfect spin-current
modulation. If we choose a typical channel in an
In$_{0.53}$Ga$_{0.47}$As semiconductor, with $m^*=0.042\,m_o$, and
spin-orbit coupling constant $\alpha=1.5 \times 10^{-11}$\,eV\,m,
\cite{Engels} this implies $W\ll 0.38\mu$m.  This width is of the
order or much smaller than the characteristics lateral widths
of the 2DEG channels patterned in the current experimental studies
of spin-injection.\cite{Hammar,Gardelis,Filip} This would suggest
that in order to achieve better  results in the spin-injection
modulation, the best choice would be ballistic quasi-one
dimensional systems (by introducing split gates for instance), where
just a few populated subbands are allowed, rather than the
quasi-two-dimensional  situation, in which many propagation channels
exist (Hall-bar experiments), with a concomitant small  subband
spacing comparable to the zero field spin-splitting energy of the
2DEG.

Recently, Moroz and Barnes\cite{Moroz-Barnes} in a theoretical study
of the effect of the spin-orbit interaction on the ballistic
conductance and the subband structure of  quasi-one-dimensional
electron systems, showed that a drastic change in the
{\it k}-dependence of the subband spectrum occurs  with respect to
the purely 2DEG system when  relatively strong spin-orbit coupling
is considered. This yields  additional subband  extrema and subband
anticrossings, as well as anomalous peaks in the conductance of the
Q1DEG.  To what extent these effects can influence the behavior of a
quasi-one-dimensional spin-modulator device  has not been
investigated, and this is the aim of the present work.

In this paper we investigate the effect of the strength of the
Rashba  spin-orbit coupling on the spin transport properties of
narrow quantum wires.  We find it convenient for this purpose to work
in a simple  tight-binding approach in which an homologous version
of the Rashba spin-orbit coupling is employed. In particular, we
find that the spin-orbit interaction induces dramatic
qualitative changes in the spin-polarized current transmitted through
Q1DEG systems, provided that a strong spin-orbit coupling is
present. A strong dependence of the spin-conductance on the incident
Fermi energy is found to be correlated with subband mixing induced
by a robust spin-orbit coupling. This dependence can significantly
suppress the spin-modulation at finite temperatures and or bias
voltages. These results may have important implications for
prospective quasi-one-dimensional spin injector devices at room
temperature or under large applied voltages.

The remainder of the article is organized as follows: In Sec.\ II the
theoretical approach is developed, starting with a brief summary of
the relevant features of the Rashba Hamiltonian and its induced
spin precession effect. A tight-binding model for the Rashba
Hamiltonian is also presented here. A brief description of the
conductance calculation is given at the end of the section. The
numerical results and main conclusions are given in Sec.\ III, and
finally the criteria that distinguish between the {\it weak} and {\it
strong} spin-orbit coupling regimes and the method
used to obtain the subband
spectrum are outlined in the Appendices.

\section{Theoretical Model}
\subsection{Hamiltonian for the Rashba effect}

In the absence of a magnetic field, the spin degeneracy of the 2DEG
energy bands at ${\bf k}\ne 0$ is lifted by the
coupling of the electron spin with its orbital motion. This coupling
arises because of the inversion asymmetry of the  potential that
confines the 2DEG system. The spin-split dispersion involves a
linear term in {\bf k}, as was first introduced by Bychkov  and
Rashba. \cite{Rashba,Bychkov-Rashba} The mechanism is popularly
referred to as the Rashba-effect. The spin-orbit (Rashba) model is
described by the Hamiltonian

\begin{equation}
H_{so}=\frac \alpha \hbar ( \vec{ \sigma }\times
\vec p)_z=i\alpha (\sigma _y\frac \partial
{\partial x}-\sigma _x\frac \partial {\partial y})
\end{equation}

Here the $z$ axis is chosen perpendicular to the 2DEG system (lying
in the $x-y$ plane),
$\alpha$ is the spin-orbit  coupling constant (Rashba parameter)
which is sample dependent, and is proportional to the interface
electric field, $\vec{\sigma}=(\sigma_x,\sigma_y,\sigma_z)$  denotes
the spin Pauli matrices,  and $ {\vec p}$ is the momentum
operator. The
experimental values of $\alpha$ for different materials  range from
about
$6\times 10^{-12}$ eV\,m at electron densities of
$n=0.7\times 10^{12}\, $cm$^{-2}$  to $3.0 \times 10^{-11}$\,eV\,m
at  electron densities of $n=2\times 10^ {12}\,$cm$^{-2}$.
\cite{Nitta,Heida,Engels,Matsuyama}

The Rashba Hamiltonian (1), which is derivable from group
theoretical arguments\cite{Winkler}, is invariant under time
reversal, that is
$[\hat T, H_{so}]=0$. The time reversal operator is represented
here by $\hat T = i\sigma_y \hat C$, with $\hat C$ the complex
conjugation operator. Since the degeneracy of the electronic states
at ${\bf k}=0$ can be only lifted if the time reversal symmetry of
the system is broken, it follows that the Rashba Hamiltonian, (due
to its time reversal invariance), can not produce a spontaneous spin
polarization of the electron states. Nevertheless, as mentioned
earlier, it is capable of removing the spin degeneracy for
${\bf k}\ne 0$. This is made clear by noticing that the total
effective mass Hamiltonian for a 2DEG system, as a result of the
Rashba effect [Eq. (1)],  has the form

\begin{equation}
H=\left(
\begin{array}{cc}
\frac{\hbar^2}{2m^*} (k_x^2 +k^2_y)   & i\alpha k_x + \alpha k_y \\
  -i\alpha k_x + \alpha k_y &  \frac{\hbar^2}{2m^*} (k_x^2 +k^2_y)
\end{array}
\right),
\end{equation}

\noindent where $H=H_o+H_{so}$, with $H_o$ being  the electronic
kinetic energy part in the absence of the Rashba effect.
Clearly, the Hamiltonian  $H$ produces two separate branches
for the electron states,

\begin{equation}
E(k) = \frac{\hbar^2}{2m^*}k^2 \pm\alpha k,
\end{equation}

\noindent here $k=|{\bf k}|$ is the magnitude of the 2D wavevector
  in the 2DEG plane.
Since the spin-orbit coupling  $\alpha$ depends on the interface
electric field, it is possible to tune the strength of the
splitting between the two branches by applying an external gate
voltage,  which will alters the net effective electric field at
the interface, as has been verified in several
experiments.\cite{Nitta,Grundler,Matsuyama}

\subsection{The Rashba spin-precession effect}

The Rashba effect is the basis for the proposed, and yet to be
implementated, Datta and Das spin-modulator device.\cite{Datta-Das}
In this device, a spin-polarized current is injected from a
ferromagnetic material into a 2DEG at a inversion layer (formed at a
semiconductor heterojunction) and then collected by a second
ferromagnetic material (Fig. 1(a)). In basic terms, the idea is that
the Rashba  effect will induce a spin-precession of the electrons
moving parallel to the interface, rotating them
with respect to the  magnetization direction of the second
ferromagnet (collector). Then by adding a gate voltage the net
effective electric field (and hence, the spin-orbit
interaction) at the interface can be modified,  tuning  the
spin precession, and   therefore, the transmitted
spin-polarized current is modulated
accordingly.\cite{Datta-Das}

We mentioned in the introduction that an important prerequisite to
having an overall  strong spin current modulation that the
angular distribution of the 2DEG be restricted by imposing a
transverse confining potential. Bearing this in mind, we
will now summarize the essential physics of the
spin-precession effect in a Q1DEG system for the case of
{\it weak} spin-orbit coupling. The summary will define the
basic conceptual framework that will be needed to understand
our results in the general multichannel case with arbitrary
spin-orbit coupling strength and will also establish the
notation to be used in the remainder of the paper.

Consider a Q1DEG system which is defined by  applying
split gates to a 2DEG in a semiconductor heterostructure
(In$_x$Ga$_{1-x}$As/In$_y$Al$_{1-y}$As for instance, see Fig. 1(b)).
Due to the confining potential
$V(x)$, the electron motion will be quantized in the $x$ direction,
Fig. 1(c). Following Datta and Das\cite{Datta-Das}, let's
assume that the Rashba spin-orbit interaction $H_{so}$ is
sufficiently weak that its effect can be incorporated
pertubatively. For such a case, the  unperturbed ($\alpha=0$)
Hamiltonian will satisfy
$H_o|n,\sigma\rangle=E^o_n|n,\sigma\rangle$, where the eigenvalues
are given by $E^o_n=E_n +\hbar^2k_y^2/2m^*$, with $n$ denoting the
subband index.  The unperturbed spin degenerate  eigenstates
have the form, $|n,\sigma\rangle \rightarrow e^{ik_y
y}\phi_n(x)|\sigma\rangle$, where $\sigma
=\uparrow,\downarrow$  with the definitions of the spinors
$|\uparrow\rangle={\tiny
\left(
\begin{array}{c}
  1  \\
  0
\end{array}
\right)
}
$, and
$|\downarrow\rangle={\tiny
\left(
\begin{array}{c}
  0  \\
  1
\end{array}
\right)
}
$. Note that the $\phi_n(x)$ are the solutions of
\begin{equation}
\left(-\frac{\hbar^2}{2m^*}\frac{d^2}{dx^2}+V(x)\right)\phi_n(x)=E_n\phi_n(x).
\end{equation}

We seek  the eigenvalues for the system in which
we have a non-zero and weak spin-orbit interaction, $\alpha\ne
0$ ({\it i.e.} $H_{so}\ne 0$). That is for $H=H_o+H_{so}$. From
degenerate perturbation theory, we obtain  (to zeroth oder) the
following system of equations for each subband
$n$,
\begin{equation}
(E^o_n +
(H_{so})^{\sigma\sigma}_{nn}-E)a^{o}_{n\sigma}+
\sum_{\sigma'\ne\sigma}(H_{so})^{\sigma\sigma'}_{nn}a^{o}_{n\sigma'}=0
\end{equation}

\noindent with
$a^{o}_{n\sigma}$ the zeroth order expansion for the coefficients
  $a_{n\sigma}$ used to expand the perturbed states in terms of the
known unperturbed states $|n,\sigma\rangle$. The result in Eq. (5)
is valid as long the condition

\begin{equation}
\left|\frac{(H_{so})^{\sigma\sigma'}_{nm}}{(E^o_m-E^o_n)}\right|\ll1,
\end{equation}

\noindent for $m\ne n$ has been fulfilled, where
$(H_{so})^{\sigma\sigma'}_{nm} =\langle
n,\sigma|H_{so}|m,\sigma'\rangle$ are the matrix elements that
intermix the different subbands and spin-states in the perturbed
system. Explicitly  they are given by

\begin{eqnarray}
(H_{so})^{\uparrow\downarrow}_{nm} & = &\alpha k_y
\delta_{nm}+\alpha\langle n|\frac{d}{dx}|m\rangle  \\
(H_{so})^{\downarrow\uparrow}_{nm} & = &\alpha k_y
\delta_{nm}-\alpha\langle n|\frac{d}{dx}|m\rangle  \\
(H_{so})^{\uparrow\uparrow}_{nm} & = &
(H_{so})^{\downarrow\downarrow}_{nm} =0.
\end{eqnarray}

Clearly the perturbation is non-diagonal in the spin index and it is
linear in $k_y$. If we suppose that the transverse confining
potential $V(x)$ has reflection symmetry in $x$,
eqs. (7) and (8) will reduce to
\begin{equation}
(H_{so})^{\uparrow\downarrow}_{nn} =
(H_{so})^{\downarrow\uparrow}_{nn} = \alpha k_y.
\end{equation}

\noindent Inserting the results from (9) and (10) into equation (5)
we get for each channel $n$,

\begin{equation}
\left(
\begin{array}{cc}
E^o_n - E & \alpha k_y \\
\alpha k_y & E^o_n - E
\end{array}
\right)
\left(
\begin{array}{c}
a^o_{n\uparrow} \\
a^o_{n\downarrow}
\end{array}
\right)=0.
\end{equation}
which yields the eigenvalues

\begin{equation}
E^{\pm}(k_y)=E_n+\frac{\hbar^2}{2m^*}k_y^2\pm\alpha k_y.
\end{equation}

Eq. (12) shows that for Q1DEG systems, the Rashba spin-orbit
interaction introduces (to zeroth order), a lifting of the spin
degeneracy for each subband state $n$. The nature of the
splitting is such that it allows  electrons with the same energy
to  have different wave vectors, ($k_{y1}$ and $k_{y2}$), that
is
$E^{+}(k_{y1})=E^{-}(k_{y2})$,   where
$k_{y1}$ is the wave vector associated with the subband $E^{+}$
with eigenvector {\tiny $
\left(
\begin{array}{c}
  1  \\
  1
\end{array}
\right)
$
},
whereas $k_{y2}$ represents the wave vector associated with the
subband $E^{-}$ with eigenvector  {\tiny $
\left(
\begin{array}{c}
  1  \\
  -1
\end{array}
\right)
$
}. Therefore, if we were to drive spin-up polarized
electrons  into the Q1DEG using a ferromagnetic material as in the
Datta and Das device (Fig. 1(a)), the wave emerging from
the semiconductor wire  would be represented as
$\psi =${\tiny $
\left(
\begin{array}{c}
  1  \\
  1
\end{array}
\right)
$
}
$
e^{ik_{y1}L}
+ 
{\tiny
\left(
\begin{array}{c}
  1  \\
  -1
\end{array}
\right)
}
e^{ik_{y2}L}
$,
where $L$ is the length of the semiconductor wire.
Therefore the probability of detecting a spin up {\tiny $
\left(
\begin{array}{c}
  1  \\
  0
\end{array}
\right)
$
}
electron at the collector contact would be proportional to

\begin{equation}
|\langle (1\, 0)|\psi\rangle|^2 = 4\cos^2\frac{(k_{y2}-k_{y1})L}{2},
\end{equation}

\noindent whereas if the collector
contact is magnetized such that it detects only spin-down polarized
electrons, the probability of detecting a spin-down  ${\tiny
\left(
\array{c} 0 \\ 1 \endarray \right)}$ electron will be
proportional to

\begin{equation}
|\langle (0\, 1)|\psi\rangle|^2 = 4\sin^2\frac{(k_{y2}-k_{y1})L}{2}.
\end{equation}

The results (13) and (14) are very important; they imply that if we
inject and collect spin-polarized (up or down) electrons into (and
from) a Q1DEG system, the Rashba-effect will produce   a modulation
of the transmitted current at drain contact with a differential
phase shift given by
$\Delta\theta=\Delta  k_y\,L$, where $\Delta k_y = k_{y2}-k_{y1}$
. In other words, the Rashba effect would induce a spin precession of
the transmitted electrons with a phase shift $\Delta \theta$ with
respect to those injected at the ferromagnetic emitter.
 From Eq. (12) is straightforward to determine $\Delta k_y$. Since
$E^{+}(k_{y1})=E^{-}(k_{y2})$ we have that

\begin{eqnarray}
E^{+}(k_{y1})-E^{-}(k_{y2}) & =
&\frac{\hbar^2}{2m^*}(k_{y1}^2-k_{y2}^2)+
\alpha(k_{y1}+k_{y2}) \nonumber \\
& = & 0
\end{eqnarray}

\noindent which leads to  $k_{y2}-k_{y1}= 2m^*\alpha/\hbar^2$.
Therefore the differential phase shift $\Delta \theta$ can  be
written  as\cite{Datta-Das},

\begin{equation}
\Delta\theta=\frac{2m^*}{\hbar^2}\alpha L,
\end{equation}

\noindent which is proportional to strength of the Rashba
parameter $\alpha$  and to the separation {\it L} between the
magnetic contacts or the length of the Q1DEG system. Then by
applying a back gate voltage bias, the Rashba parameter can be
varied in principle, and hence, the degree of the electron spin
precession would be tuned correspondingly.

Eq. (16) suggests that spin-current modulation can be attained in
semiconductor nanowires, regardless of the mode number and large
applied bias, as indeed Datta and Das concluded.\cite{Datta-Das}
Notice as well that the  phase shift is  energy independent.
We should recall here that these results were obtained with the
premise that the Rashba spin-orbit interaction was weak enough,
such that  the confinement energy was much larger than the
spin-splitting energy induced by the Rashba effect, and therefore,
the intersubband mixing was neglected. In other words, whenever the
condition (6) is satisfied. For the hard wall confining potential
considered here this means

\begin{equation}
\left|\frac{(H_{so})^{\sigma\sigma'}_{nm}}{(E^o_m-E^o_n)}\right|\approx
\frac{\alpha m^* W}{\hbar^2}\ll 1
\end{equation}

\noindent which would imply  $W\ll\hbar^2/\alpha m^*$ and gives us
a rough upper limit of the width of the confinement for the
intersubband mixing to be neglected. Typical values for $m^*$ and
$\alpha$ yield widths of $W\ll 0.38\,\mu$m. We will arrive at the
similar conclusions using parabolic potential for the transverse
confinement
$V(x)$.

The criterion (17) for the validity of the pertubative theory
discussed above is quite restrictive since it requires {\it both}
weak spin-orbit coupling and a narrow wire. Thus in general the
intersubband mixing that has been neglected in the above discussion
can be important, and the simple energy-independent expression (16)
for the differential phase shift may then no longer apply. This
results in important qualitative changes in the electron
spin precession and in the behavior of the spin-modulation devices
as will we discuss below.

In the following section we present a tight-binding analog of the
Rashba Hamiltonian that can be used to study spin transport in cases
where intersubband coupling is important and that pertubative
theory fails, and also in more general geometries which do not
lead themselves readily to analytic solution. We then proceed to
solve the multi-channel scattering and spin-dependent transport
problem exactly (numerically) within the tight-binding formalism in
a variety of situations where intersubband mixing is significant
and needs to be considered.

\subsection{Rashba Hamiltonian: A tight binding model}

In this section, the spin-orbit interaction given by the Rashba
Hamiltonian is reformulated within the tight-binding approach in a
lattice model. We consider a quasi-one dimensional wire, which
is assumed to be infinitely long in  the propagation direction. The
wire is represented by a two dimensional grid  with lattice constant
$a$. We choose the coordinate system such that the $x$  axis, with
$N_x$ lattice sites, is in the transverse direction, while the $y$
axis, with $N_y$  lattice sites is in the longitudinal direction,
Fig. 2(a).

We assume only nearest neighbor spin-dependent interactions for the
Rashba perturbation.  For the purpose
of the present model, the localized site orbitals  will be assumed to
have the symmetry of $s$-states. Then the tight-binding analog of
eq. (1) takes the form

\begin{eqnarray}
\widehat{H}_{so}^{tb} = & - &t_{so}\sum_{\sigma ,\sigma ^{\prime }}
\sum_{l,m}\Big( C_{l+1,m,\sigma ^{\prime }}^{\dagger }(i\sigma
_y)_{\sigma ,\sigma ^{\prime }}C_{l,m,\sigma} \nonumber \\
  & - & C_{l,m+1,\sigma ^{\prime }}^{\dagger }(i\sigma _x)_{\sigma ,
\sigma ^{\prime}}C_{l,m,\sigma }\Big) + H.c.
\end{eqnarray}

\noindent with an isotropic nearest-neighbor transfer integral
$t_{so}$ which measures the strength of the Rashba
spin-orbit interaction and will be shown below to have the value
$t_{so}=\alpha /2a$, $C_{l,m,\sigma }^{\dagger }$ represents the
electron creation operator at site
$(l,m)$ with spin state
$\sigma$, ($\sigma =\uparrow,\downarrow$). Here $\sigma _x$
and $\sigma _y$ are the Pauli spin  matrices. Notice that
$t_{so}>0$, and in principle it can be tuned by an external electric
field.

This Hamiltonian is formally similar to that studied previously by
Ando and Tamura\cite{Ando-Tamura} in the context of conductance
fluctuations and localization in  quantum wires with
boundary-roughness scattering. We emphasize that in the present
model no roughness scattering is present, and that here the physical
origin of spin-orbit scattering is the
asymmetry of the electric field in the quantum well that  contains
the 2DEG.

We divide the wire  into three main regions. In two of these (I and
III in Fig. 2(a))  which are near the ferromagnetic
source and drain,  the spin-orbit
hopping parameter $t_{so}$ is set to zero. We assume that the
semiconductor interface at which the Rashba effect occurs does not
extend into these two regions and that $t_{so}=0$ there for that
reason. In the middle region (II) the spin-orbit coupling is
finite, ($t_{so}\ne 0$) at the semiconductor interface. In the
actual calculation this region (II) is further divided into three, to
introduce  the spin-orbit interaction adiabatically. Therefore, the full
Hamiltonian reads

\begin{equation}
\widehat{H}=\widehat{H}_o+\widehat{V}+\widehat{H}^{tb}_{so},
\end{equation}

\noindent where the spin diagonal parts of $\hat{H}$ are given by

\begin{eqnarray}
\widehat{H}_o & = & \sum_{l,m,\sigma}\epsilon\,
C_{l,m,\sigma }^{\dagger}C_{l,m,\sigma }+ \nonumber \\
& + & t \sum_{l,m,\sigma}\left( C_{l+1,m,\sigma }^{\dagger
}C_{l,m,\sigma }+C_{l,m+1,\sigma }^{\dagger }C_{l,m,\sigma }
+H.c.\right), \nonumber \\
\end{eqnarray}

and

\begin{equation}
\widehat V = \sum_{l,m,\sigma}V_{l,m}C_{l,m,\sigma
}^{\dagger}C_{l,m,\sigma}\, ,
\end{equation}

\noindent with $\epsilon $ the on-site energy,
$t$ is the hopping energy ($t=-\hbar^2/2m^*a^2$), and $V_{l,m}$ is an
additional confining potential. The full Hamiltonian $\widehat{H}$
will have eigenvectors given by

\begin{equation}
|\Psi \rangle =\sum_\sigma \sum_{l,m}\chi _{l,m}^\sigma
|l\ m\ \sigma\rangle ,
\end{equation}


\noindent where we have defined the spinors,

\begin{equation}
\Psi_{l,m}^{\uparrow}=\chi_{l,m}^{\uparrow}\left(
\begin{array}{c}
  1  \\
  0
\end{array}
\right) ,\,\text{and}\,\Psi_{l,m}^{\downarrow}=
\chi_{l,m}^{\downarrow}\left(
\begin{array}{c}
0 \\
1
\end{array}
\right)
\end{equation}

\noindent with $\chi _{l,m}^{\sigma}=\langle l\,
m\,\sigma|\Psi\rangle$, $\chi _{l,m}^{\sigma}$ being the
electronic wave function at site $(l,m)$ and in spin state
$\sigma=\uparrow,\downarrow$. In (22), $|l\ m\ \sigma \rangle
=C_{l,m,\sigma }^{\dagger }|0\rangle $, with $|0\rangle$ denoting
the vacuum, and we assume
$\langle l^{\prime }\ m^{\prime }\ \sigma ^{\prime }|l\ m\ \sigma
\rangle =\delta _{l,l^{\prime }}\delta _{m,m^{\prime }}
\delta _{\sigma ,\sigma^{\prime }}$.

We establish the correspondence between Eqs. (18) and (1) and
determine the value of $t_{so}$ by noticing that for a
2DEG system with  plane wave solutions  of the form
$e^{i(k_xal+k_yam)}$, the Hamiltonian $\widehat{H}$ yields
the two-dimensional tight-binding eigenvalues,

\begin{equation}
E({\bf k}_\parallel)=E_o({\bf k}_\parallel)\pm 2t_{so}
\sqrt{\sin {}^2(k_xa)+\sin
{}^2(k_ya)}
\end{equation}

\noindent with $E_o({\bf k}_\parallel)=\epsilon+2t[\cos (k_xa)+
\cos (k_ya)]$ the tight-binding conventional subband dispersion for
a 2DEG, with ${\bf k}_\parallel=(k_x,k_y,0)$. For small
$k_x a$ and $k_y a$ (setting the on-site energy $\epsilon$ equal to
$-4t$), Eq.\ (24) reduces to

\begin{equation}
E=\frac{\hbar ^2}{2m^{*}}(k_x^2+k_y^2)\pm 2at_{so}\sqrt{k_x^2+k_y^2}
\end{equation}

\noindent  Note that
Eq. (25) is just the continuum subband dispersion (3), and will
correspond to the expected Rashba subband linear splitting with the
definition $\alpha=2at_{so}$.
\cite{tso-values}

\subsection{Spin transport calculation}

For the study of the spin-dependent transport in a Q1DEG system, the
physical model we bear in mind is as follows: Consider a Q1DEG
spin-modulator device as shown in Fig. 2(b). The device has two
independent ferromagnetic source contacts with magnetizations
oriented  such that one of them  can emit only spin-up
polarized electrons (contact SF$^{(\uparrow)}$), whereas the second
source contact SF$^{(\downarrow)}$, can emit only spin-down polarized
electrons. Likewise, two independent ferromagnetic drain
contacts DF$^{(\uparrow)}$ and DF$^{(\downarrow)}$ are attached at
the opposite end of the device which are able to detect just
spin-up and spin-down polarized electrons, respectively. A perfect
ohmic contact between the ferromagnetic materials and the
semiconductor is assumed. A back gate underneath the device and
directly below the Q1DEG channel will control the spin
precession (through the Rashba effect) of the injected electrons. We
will suppose that the device is set up such that spin polarized
electrons are launched either from the SF$^{(\uparrow)}$ or
SF$^{(\downarrow)}$ contacts, while both spin orientations
($\uparrow$ and $\downarrow$) are being absorbed at the drain
contacts, ensuring in this way a spin-resolved measurement. In
addition it is supposed that the electron spin
reorientation due to defect scattering can be neglected, {\it
i.e.}, the spin relaxation time is much longer than the electron
dwell time in the device.

We now discuss the approach used in this work to study
spin-transport in the Q1DEG system described above,
specifically, for the calculation of the spin-conductance.
The spin-transport problem was solved numerically through
the use of the spin-dependent  Lippman-Schwinger equation,

\begin{equation}
|\Psi\rangle = |\Phi\rangle + G_o(E)\widehat{U}|\Psi\rangle
\end{equation}


\noindent where $|\Phi\rangle$ is the unperturbed wave function,
{\it i.e.} an  eigenstate for $\widehat H_o$, while
$G_o(E)=(E+i\epsilon-\widehat H_o)^{-1}$ is the Green's function for
the system in the absence of any kind of scattering. Here
$\widehat{U}= \widehat{V} + \widehat{H}_{so}$, represents  the
scattering part, with  $\widehat{V}$ the scattering due to the
confinement potential, and
$\widehat{H}_{so}$ the spin-dependent part due to the Rashba coupling
in the semiconductor wire. The unperturbed functions, $|\Phi\rangle$
and $G_o(E)$ are known analytically, see for instance Nonoyama {\it
et al.}\cite{Nonoyama} Notice that the unperturbed Green's  function
is also diagonal in spin index, that is

\begin{equation}
\langle l_qm_q\sigma _q|G_o|l\,m\,\sigma \rangle =
\langle l_qm_q|G_o|l\,m \rangle \delta _{\sigma \sigma _q},
\end{equation}

\noindent therefore, for a wave function at any site $(l_q,m_q)$ and
state $\sigma_q$, the Lippman-Schwinger equation for this system can
be rewritten as follows

\begin{eqnarray}
  \Psi_{l_q,m_q}^{\sigma_q} & = &  \Phi_{l_q,m_q}^{\sigma_q}
+\sum_{l,m,\sigma}\Big [  \langle l_qm_q|G_o|l\ m\rangle V_{lm}
\Psi_{l,m}^{\sigma} \,\delta_{\sigma,\sigma_{q}}
\nonumber \\
  & - & t_{so} \langle l_qm_q|G_o|l+1\ m\rangle(i\sigma_y)_{\sigma,
\sigma _q} \Psi_{l,m}^{\sigma}\nonumber \\ & + & t_{so}\langle
l_qm_q|G_o|l\ m+1\rangle (i\sigma _x)_{\sigma ,\sigma _q}
\Psi_{l,m}^{\sigma} \nonumber \\ & - & t_{so}\langle l_qm_q|G_o|l\
m\rangle (i\sigma_y)_{\sigma _q,\sigma }^{*}
\Psi_{l+1,m}^{\sigma} \\
& + & t_{so}\langle l_qm_q|G_o|l\ m\rangle(i\sigma _x)_{\sigma _q,
\sigma }^{*}\Psi_{l,m+1}^{\sigma}\Big ].
\nonumber
\end{eqnarray}

By solving the coupled linear equations resulting from the  equation
above, it is possible to determine the perturbed wave function
associated with the complete Hamiltonian $\widehat H$ at any  site of
interest.

Once the wave functions for each state $\sigma$ are known at the
ferromagnetic contacts (regions I and III in Fig. 2(a)) the
spin-dependent conductance is obtained within  the
Landauer framework
\cite{Landauer}, which is appropriate if the
electron-electron interactions are unimportant.\cite{Hausler} In
our calculations, only spin-up polarized electrons  are injected
from the emitter (region I in Fig. 2(a)) into the spin-orbit
interaction region (where the spin precession of the incident
electron is induced). At the collector region, the  transmitted
current is calculated separately for each spin, modeling the pair
of drain contacts DF$^{(\uparrow)}$ and DF$^{(\downarrow)}$ in Fig.
2(b). Therefore, in general we will have two contributions for the
net transmitted current, one coming from  spin-up electrons
that arrive at the collector (region III), and the
other one from the collected spin-down electrons.
We therefore define the conductances $G^{\uparrow}$ and
$G^{\downarrow}$ associated with the currents flowing between
SF$^{(\uparrow)}$ and DF$^{(\uparrow)}$, and DF$^{(\downarrow)}$,
respectively, by

\begin{equation}
G^{\uparrow}(E) =
\frac{e^2}{h}\sum_\upsilon \tau _\upsilon ^{\uparrow},
\end{equation}

and

\begin{equation}
  G^{\downarrow}(E)=
\frac{e^2}{h}\sum_\upsilon  \tau
_\upsilon ^{\downarrow}.
\end{equation}

It is important to note that the spin-down partial conductance
at the collector ferromagnet $G^{\downarrow}$ arises due to the
induced spin precession of the incident spin-up polarized electrons
and since no spin-down  polarized electrons are being injected.
In Eq. (29)
$\tau _\upsilon ^{\uparrow}$  is the partial transmission probability
(summed over the incident channels) that an incident electron
(injected from SF$^{(\uparrow)})$ with spin $\sigma =\uparrow,$  is
transmitted to the right ferromagnet DF$^{(\uparrow)}$,
$\upsilon
$ denotes the outgoing transmitted mode, while
$\tau _\upsilon ^{\downarrow}$ is the partial transmission
probability that an incident electron with the same spin $\sigma
=\uparrow$, is transmitted as a spin-down electron, and measured
at DF$^{(\downarrow)}$ .  The partial transmission probability is
given explicitly by

\begin{equation}
\tau _\upsilon ^\sigma =\sum_\mu \left(
\frac{v_\upsilon^\sigma }{v_{\mu}^\sigma }
\right)|t_{\upsilon \mu}^\sigma |^2
\end{equation}

\noindent where $v_\upsilon^\sigma$ and $v_\mu^\sigma$ are the
outgoing and incident electron velocities at the Fermi energy with
spin $\sigma$ and modes $\upsilon$ and $\mu$ of the drain and source
contacts, respectively. In (31) $t_{\upsilon\mu}^\uparrow$ is the
partial transmission amplitude that an incident electron with spin
$\sigma=\uparrow$ and mode $\mu$ is transmitted in the $\upsilon$
mode with the same spin state $\sigma=\uparrow$, and
$t_{\upsilon\mu}^\downarrow$ is the partial transmission amplitude
that an incident electron with spin $\sigma=\uparrow$ and mode $\mu$
is transmitted in the $\upsilon$ mode with the opposite spin state,
{\it i.e.} $\sigma=\downarrow$\,.

\section{NUMERICAL RESULTS AND THEIR IMPLICATIONS}

In order to study the dependence of the spin-conductance on the
strength of the spin-orbit interaction, we will consider two cases,
namely, {\it weak} and {\it strong}  coupling. The criteria that
distinguish these two cases are obtained in the appendix A; the
physical considerations are essentially as follows:
Since in the  multichannel scattering process the eigenstates of the
full Hamiltonian  are (in general) linear combinations of the different
spin-subbands (due to the Rashba-term), therefore, in a pertubative
sense, the contribution of the mixing of the spin-subbands should be
negligible as long the subband spacing (separation in energy)
$\Delta E_W=E_m-E_n$, is much greater than the subband
intermixing energy, Eq. (6).
  However, if the confinement energy and/or the spin-orbit coupling
are of the same order as the energy shift introduced by the intersubband
mixing  contribution, then the above condition no longer
holds. For this case we find (within a two-subband model) a
critical value for the spin-orbit coupling (see Appendix A), which
is given by     $\beta_{so} \approx (\pi a/W)^2/[(\pi a/W)+
ak_F]=\beta_{so}^{c}$, where
$\beta_{so}=t_{so}/|t|$, and $k_F$ is the Fermi wave number. The
critical value $\beta_{so}$  will therefore
define a {\it weak} spin-orbit  coupling regime whenever
$\beta_{so}<\beta_{so}^{c}$, and {\it strong} coupling regime if
$\beta_{so}> \beta_{so}^{c}$.
It should be noted that the critical
value $\alpha^{c} = \pi^2 \hbar^2 /[m^* W(\pi+Wk_F)]$ of the Rashba
spin orbit coupling parameter $\alpha$ that corresponds to $\beta_{so}^{c}$
and therefore separates the weak and strong coupling regimes
depends on the width $W$ of the wire.

In all our calculations, unless
otherwise stated, the Rashba spin-orbit interaction is turned on
and off adiabatically through a cosine-like function, with $\ell$
being the length of the adiabatic region in lattice sites. For the
remainder of the paper we will work in units of $|t|$ for all the
energies, hence we  will refer to $\beta_{so}$ and $t_{so}$
interchangeably. It is remarked that the calculations presented
here are for narrow wires, in which the quantization in the
transverse direction to the propagation is crucial.


In Fig. 3 we show the ballistic conductance for four strength
values of the spin-orbit hopping parameter $t_{so}$ for a narrow
wire of width $W=60$ nm and length $L=150$ nm. Here as in the
rest of the numerical results and as discussed in Sec.\ II, purely
spin-up polarized electrons are injected into the wire. Only the
first mode is shown for clarity. In Fig.  3(a) the case
$t_{so}=0$  is plotted and  since there is no
electron spin-precession, all  the transmitted electrons are spin-up
polarized.  When the spin-orbit parameter
is increased to $0.03$, Fig. 3(b), the precession effects  become
evident, about the $70\%$  of the detected conductance is due to
spin-up electrons,  while about $30\%$ is due to spin  down
electrons. For $t_{so}=0.08$, Fig. 3(c), the spin  conductance is
now reversed with respect to (a), that is, the net detected
conductance is due  only to spin-down electrons,
$G^{\uparrow}=0$ and $G^{\downarrow}=1.0$ (in units of $e^2/h$).
Increasing the  spin-orbit parameter further to $0.1$, the spin-up
conductance
$G^{\downarrow}
\neq 0$  once again and the spin-down conductance $G^{\uparrow}<1.0$,
Fig. 3(d).  Notice that in all cases, the spin conductance is almost
independent of the incident Fermi energy; the situation is more
complex when strong coupling is assumed, as we shall see later. The
qualitative behavior  of the spin conductance described above  is
consistent (to a good approximation) with
the energy-independent electron spin modulation predicted by Datta
and Das.\cite{Datta-Das}

The spin conductance modulation is seen clearly when we plot $G^{
\uparrow\downarrow}$ against the spin-orbit hopping parameter
$t_{so}$, see Fig. 4(a). Here the incident Fermi energy was fixed to
$0.5$ $(k_F\approx 0.7a^{-1})$ and $W=6a=60$\,nm, which gives a
critical value
$\beta_{so}= 0.22$. This value of $\beta_{so}$ separates the
sinusoidal behavior of $G^{\uparrow\downarrow}$ (predicted by Eqs.
(13), (14) and (16)) for
$\beta_{so}\le 0.22$ from its behavior for $\beta_{so}>0.22$ (the
strong spin subband  mixing regime) where the confinement energy is
of the order of the intersubband mixing energy. The effect is
clearer for a wire with
$W=120\,$ nm, [see Fig. 4(c)] for which the critical value of
$\beta_{so}$ is $0.07$.

The importance of the intersubband mixing  contribution can be
illustrated as follows. Let us write the Rashba Hamiltonian as the
sum of two terms, $H_{so}=H_{so}^{(x)} + H_{so}^{(y)}$, with
$H_{so}^{(x)}=i\alpha_x\sigma_y \partial/\partial x$ and
$H_{so}^{(y)}=-i\alpha_y\sigma_x \partial/\partial y$, where in
general $\alpha_x=\alpha_y=\alpha$ for a 2DEG. Therefore, Eqs. (7)
and (8) would be rewritten as  $(H_{so})^{\uparrow\downarrow}_{nm}
= \alpha_{y} k_y \delta_{nm}+\alpha_{x}\langle
n|\frac{d}{dx}|m\rangle$, and
$(H_{so})^{\downarrow\uparrow}_{nm} = \alpha_{y} k_y
\delta_{nm}-\alpha_{x}\langle n|\frac{d}{dx}|m\rangle$, respectively.
Now, since the only  matrix elements that incorporate the mixing of
the different subbands  are given by
$\langle n,\sigma |H_{so}^{(x)}|m,\sigma'\rangle$, with
$n\ne m$ and $\sigma \ne \sigma'$, therefore by setting $\alpha_x$
to zero (with $\alpha_y\ne 0$), the contribution of these matrix
elements can be fully suppressed. This situation is shown in Fig.
4(b) and 4(d). Note that
$G^{\uparrow\downarrow}$ are very similar in Fig. 4(a) and 4(b) for
$t_{so}<0.22$ ($t_{so}<0.07$, case 4(c)) but not for $t_{so}>0.22$
($t_{so}>0.07$, case 4(c)). Although setting $\alpha_x=0$ appears to
be  rather unphysical, it shows very explicitly, that the deviation
from the sinusoidal modulation of $G^{\uparrow\downarrow}$ for
$t_{so}>0.07$ is owed essentially, to intersubband mixing
induced by the Rashba spin-orbit coupling.

The subband dispersion for $t_{so}=0.08$ is depicted in Fig. 5(a).
The dispersion is calculated using  the procedure described in the
Appendix B, the rest of the simulation  parameters are the same  as
in Fig. 3. For this case, a linearly Rashba spin-split subband is
obtained as expected , giving for this particular case  $\Delta k =
0.157\,a^{-1}$.   The phase shift is then $\Delta \theta = \Delta k
L = \pi$, where we have used the effective  length of $L=20\,a$.
Using now formula (16),  we  obtain for $\beta_{so}=0.08$ a phase
shift $\Delta \theta = 3.2$, just slightly above of  what we obtain
from the subband dispersion calculation.\cite{phase} The
corresponding spin conductance is dotted in Fig. 5(b). Note that
a fully reversed spin-polarization occurs for all modes. Thus
the spins precess in unison in all of the subbands and even
a relatively wide multimode quantum wire should be expected to
function as an efficient Datta-Das spin-transistor in this regime.
For the same value of $t_{so}$, Fig. 5(c)  shows the conductance
vs. the effective length $L$ where the  spin-orbit coupling is
present. The oscillations of $G^{\uparrow}$ and $G^{\downarrow}$
have a differential phase  which corresponds exactly to the expected
value for length $L=20\,a$ in Eq. (16). It is important to emphasize
that the qualitative  features seen here in the conductance  for
{\it weak} $t_{so}$ remain the same for wider wires and with a
lateral constriction in the region with spin-orbit coupling. For
example in Fig. 6(a) we have plotted the conductance for a wire
having twice the width discussed above, {\it i.e.} here $W=120$ nm,
where $t_{so} =0.05$ and with a strength for a parabolic  confining
potential of
$w_x =0.2$ (here the confinement potential is given by $w_x(x/a)^2$,
with $w_x$ in units of $|t|$) .  We note the same step-like
characteristic of the ballistic conductance. Note that here as well,
  (Fig 6(a))the modulation achieved between
$G^{\uparrow}$ and $G^{\downarrow}$ remains constant regardless of
the incident Fermi energy and number of populated subbands.


Now we turn our attention to the case  $\beta_{so} >
(\pi a/W)^2/[(\pi a/W)+ ak_F]$, {\it i.e.} strong spin-orbit
interaction, for the case with
$W=60$ nm,
$\beta_{so}\ge 0.22$. Fig. 7 displays the spin conductance for four
different values of the strength for $t_{so}$.  Here the adiabatic
region is set to $\ell=20 a$, and the rest of the parameters are as
in  Fig. 3. For sake of simplicity we focus only on the first
propagating mode. It is evident  that for this regime, the
spin-conductance behavior is markedly  different from that  with
weak spin-orbit coupling discussed previously. Here $G^{\uparrow
\downarrow}$ are both strongly energy dependent. We note that the
probability of an injected spin up electron to pass unchanged
through the spin-orbit region decreases with energy [Fig.
7(a)], while the probability of detecting a  transmitted spin down
electron increases accordingly. For $t_{so}=0.3$, Fig. 7(b), a full
polarization develops when reaching an incident energy of
$E_{F}=0.8$, although without the spin flipping. Cases (c) and (d)
are quite interesting. A clear precession of $G^{\uparrow\downarrow}$
with energy is observed, having a sinusoidal like character. For
instance, in (c), the polarization of the transmitted current  has
opposite  orientation than the polarization of the injected current
for $E_{F}= 0.65$, and is reversed again at $E_{F}= 1.0$. This
surprising result (which is due to intersubband coupling and is
therefore beyond the scope of the original theory of Datta and Das)
would suggest that for certain values of the strength of the
spin-orbit interaction, just by varying the the Fermi
energy, the device can work  as a spin ``switch'' device, while the
spin-orbit interaction is kept constant. In other words, it can
function as a spin-transistor such that the switching can be tuned
just by varying the Fermi energy rather than varying the Rashba
coupling, as in Datta and Das spin-transistor.

The subband dispersion for the case $t_{so}=0.4$ of Fig. 6(b) is
shown in Fig. 8(a). For this large value of $t_{so}$, the
spin-subband dispersion deviates considerably from the typical linear
Rashba splitting for the weak spin-orbit interaction (A similar
dispersion was obtained  recently  by Barnes and Moroz
\cite{Moroz-Barnes}). The subbands are no longer parabolic which
gives rise to a dependence of $\Delta k$ on the energy. Here $\Delta
k$ is the difference between the wave vectors associated the two
spin subbands at a given energy, $ \Delta k \rightarrow \Delta k(E) =
k_2(E) - k_1(E)$. We find for example that $\Delta k(E_{F} =0.5) =
0.6489\,a^{-1} $, whereas $\Delta k(E_{F} =1.0) = 0.4533\,a^{-1} $,
which yields $\Delta \theta(E_F=0.5) =6.2\pi$ and $\Delta
\theta(E_F=1.0) = 4.3\pi$, where we have used an effective length
$L=30\,a$. The net change in the differential phase between these two
cases is thus $\delta(\Delta \theta) \approx 2\pi$. This last result
can be independently checked by looking at the spin-resolved
conductances $G^{\uparrow}$ and $G^{\downarrow}$ in Fig. 8(b) where
the oscillations of both $G^{\uparrow}$ and $G^{\downarrow}$ exhibit
phase shifts of $\approx 2\pi$ between $E_F=0.5$ and $E_F=1.0$.  We
have also plotted the spin conductance for $E_F=1.0$ versus the
length of the spin-orbit interaction region in the wire, [Fig.
8(c)]. A wave length $\lambda$ of $13.8\,a$, is extracted from the
data. The associated wave number is $\Delta k=0.45$, value that
matches  that obtained independently from the band dispersion
calculation.

Now let us return to Fig. 6 in order to analyze the {\it
strong}
$t_{so}$ regime for a wider wire ($L=12a$). In particular  Fig.
6(b) shows an analogous case to that studied in Fig. 6(a) but
with $t_{so}=0.4$. It is evident that the strong coupling here
suppresses the plateaus for $G^{\uparrow\downarrow}$ to a
significant degree. A rather complicated structure is obtained
revealing the non-trivial nature of the subband intermixing.   We
have also calculated the relative conductance change [defined as
$\Delta G/G_o= (G^{\uparrow}-G^{\downarrow})/
(G^{\uparrow}+G^{\downarrow})$] against the spin-orbit parameter
$t_{so}$ for three different effective lengths of the wire with the
Rashba interaction ($L=4W$, $8W$, and $15W$). This is shown in Fig.\
9 for a wire with width $W=6a$ and uniform parabolic potential
strength of $w_x=1.8$ over an effective length of $10a$ and at the
Fermi energy  $E_F=2.5$ for the three cases. A beat-like pattern
is found due to the spin precession, developing nodes as the length
of the channel is increased. \cite{SdH-like} In the inset the
channel length dependence of
$\Delta G/G_o$ is plotted for a typical experimental $t_{so}$, it is
clear that for such relatively weak spin-orbit interaction the
relative conductance change has a negative slope with a change in
sign (indicative of a spin-precession) in the length range shown.
This behavior resembles that observed recently by C. M. Hu  {\it et
al.}\cite{Hu-Nitta} in their length dependence measurements for the
spin precession.

A final comment on the adiabaticity: the adiabaticity in the
spin-orbit interaction was introduced in our calculations in order
to model a smooth transition between the regions with no spin-orbit
coupling ({\it i.e} near the ferromagnetic contacts) and the region
with the finite Rashba spin-orbit coupling along the quantum wire.
For the weak coupling regime the calculations of the
spin-conductance vs. Fermi energy showed plateaus with small
oscillations in the non-adiabatic case, nevertheless the
spin-precession behavior was found to be qualitatively very similar
to that  observed in the adiabatic case. However, the strong
coupling regime showed  drastic differences. Calculations without
the adiabaticity showed a rather complicated behavior with a
significant suppression of the conductance plateaus.

In summary, we have shown that a strong Rashba spin-orbit
interaction can produce dramatic changes in the spin-resolved
transmission of spin-polarized electrons injected into ballistic
narrow wires. The effects can be very significant, and can even
suppress the expected spin-modulation, as the strong (Rashba)
coupling was found to induce a complex dependence of the
spin-precession on the incident Fermi energy of the injected
electrons. These results should be of importance for the
spin-injection (modulation) in quasi-one-dimensional devices under
large bias used to tune the Rashba effect.

\acknowledgments

This work was supported by NSERC and by the Canadian Institute for
Advanced Research. The authors thank J. Nitta for kindly providing
us with unpublished results.

\end{multicols}

\clearpage
\onecolumn

\appendix

\section{Criteria for weak and strong Rashba coupling}

In this appendix the criteria we have used to distinguish the
regimes for {\it weak} and {\it strong} Rashba spin-orbit coupling
in a Q1DEG system are deduced analytically. Consider $|\Phi\rangle$
the eigenstate solution of

\begin{equation}
H|\Phi\rangle=E|\Phi\rangle,
\end{equation}

\noindent where $H=H_o+H_{so}$  is the effective mass  Hamiltonian
for a Q1DEG in the presence of the Rashba coupling,
with $H_o|n,\sigma\rangle=E^o_n|n,\sigma\rangle$, and
$E^o_n=E_n +\hbar^2k_y^2/2m^*$,  ($n$ is the
subband index and
$\sigma=\uparrow,\downarrow$), see Sec. II. B.
We expand
$|\Phi\rangle$ in terms of the eigenstates for the system in the
absence of the Rashba effect.
That is $|\Phi\rangle=\sum_{n,\sigma}
a_{n\sigma}|n,\sigma\rangle$. Inserting this expansion into (A1) and
multiplying from the left by $\langle m,\sigma'|$ yields,

\begin{equation}
(E_n^o-E)a_{n\sigma} +
\sum_{m,\sigma'}(H_{so})^{\sigma,\sigma'}_{nm}a_{m\sigma'}=0.
\end{equation}

\noindent Now since $(H_{so})^{\sigma,\sigma'}_{nm}=0$ for all
$\sigma=\sigma'$ (see eq. (9)), therefore eq. (A2) will read

%

\begin{equation}
(E_n^o-E)a_{n\sigma} +
\sum_{\sigma'\ne\sigma}(H_{so})^{\sigma,\sigma'}_{nn}
a_{n\sigma'}+
\sum_{\sigma'\ne\sigma,m\ne
n}(H_{so})^{\sigma,\sigma'}_{nm}a_{m\sigma'}=0,
\end{equation}

\noindent the second term in (A3) which is diagonal in the
subband index will corresponds to the linear Rashba splitting,
while the third term gives rise to the spin subband
intermixing. In a two-subband model (denoted below by $n$ and
$n+1$), the system of eqs. (A3) can be rewritten as

\begin{equation}
\left(
\begin{array}{cccc}
E^o_n - E & \alpha k_y & 0 & -\alpha\Delta_{so}\\
\alpha k_y & E^o_n - E & \alpha\Delta_{so} & 0 \\
  0 & \alpha\Delta_{so} & E^o_{n+1} - E & \alpha k_y \\
-\alpha\Delta_{so} & 0 & \alpha k_y & E^o_{n+1} - E
\end{array}
\right)
\left(
\begin{array}{c}
a_{n\uparrow} \\
a_{n\downarrow} \\
a_{n+1\uparrow} \\
a_{n+1\downarrow}
\end{array}
\right)=0 \, ,
\end{equation}

\noindent with the definition of the intersubband mixing term,
$\Delta_{so}=\langle n+1|\frac{d}{dx}|n\rangle$. The eigenvalues
of (A4) are explicitly,

\begin{equation}
E_{1,2}=\frac{1}{2}(E_{n+1}^o+E_n^o) \pm \frac{1}{2} \sqrt{(\Delta
E_W-2\alpha k_y)^2 +(2\alpha\Delta_{so})^2},
\end{equation}

\noindent and
\begin{equation}
E_{3,4}=\frac{1}{2}(E_{n+1}^o+E_n^o) \pm \frac{1}{2} \sqrt{(\Delta
E_W+2\alpha k_y)^2 +(2\alpha\Delta_{so})^2}
\end{equation}

\noindent with $\Delta E_W = E_{n+1}^o-E_n^o=E_{n+1}-E_n$.
Notice that for  small $\alpha\Delta_{so}$, after a
zeroth order  Taylor-series expansion of the radicals in (A5) and
(A6), the eqs. will reduce exactly to that given by
eq. (12) for the energy subband linear splitting. Such
expansion will hold as long the condition

\begin{equation}
   \frac{ (2 \alpha \Delta_{so})^2 }{ (\Delta E_W \pm 2
\alpha k_y)^2} \ll 1 \, ,
\end{equation}

\noindent is satisfied. Thus, Eq. (A7) will lead us to a
criteria for which the subband intermixing is, on the other hand,
not neglectable. For such case, explicitly

\begin{equation}
  \left |\frac{ 2\alpha \Delta_{so}}{ \Delta E_W \pm 2
\alpha k_y}   \right |\approx 1 \,.
\end{equation}

   Modeling the potential that defines the Q1DEG as a hard
well confining potential, we have $\Delta E_W
=E_{n+1}^o-E_n^o=(2n+1)(\frac{a\pi}{W})|t|$, and
$\Delta_{so}=\frac{n(n+1)}{2n+1}(\frac{4}{W})$. Inserting this
expressions in (A8) and replacing $ky\rightarrow k_F$, where
$k_F$ is the Fermi wave number, we get

\begin{equation}
\alpha \approx \left |\frac{(2n+1)^2(\pi a)^2 |t|}{8n(n+1)W
\pm2(2n+1)k_FW^2  }\right |\, ,
\end{equation}

\noindent therefore, for the first subband ($n=1$), and using
$\alpha = 2a\beta_{so}|t|$,  we get (choosing the $+$ solution in
order to get the lowest value for the ratio in (A9))

\begin{equation}
\beta_{so} \approx \frac{(\pi a/W)^2}{(\pi a/W) + ak_F}.
\end{equation}

Eq. (A10) defines the critical value of the spin-orbit
coupling at which the intersubband mixing becomes relevant.
Therefore we can define a regime where the spin-orbit coupling
is {\it weak}, whenever $\beta_{so} < (\pi a/W)^2/[(\pi a/W)
+ ak_F]$,  and a regime where the spin-orbit coupling is {\it
strong}, such that spin subband intermixing becomes important,
that is whenever $\beta_{so}> (\pi a/W)^2/[(\pi a/W) + ak_F]$.


\section{Subband spectrum calculation}

In this appendix we describe the transfer matrix method for the
calculation of the subband structure of an uniform quantum wire with
a Rashba spin-orbit interaction.  The method is a generalization
of the spin-independent Ando calculation.\cite{Ando}   The
dispersion is calculated within the lattice model for a wire with a
finite and uniform spin-orbit interaction and a lateral confinement
potential.

By applying  the states $\langle l m \sigma |$ and $|\Psi \rangle$
to the left and right of Eq. (19), respectively, we arrive at the
following discrete coupled equations [for clarity, the notation
$\sigma =\pm$, is used instead of the arrow symbols to denote spin-up
and spin-down states]

\begin{equation}
(W_{l,m}^{\pm }-E)\chi _{l,m}^{\pm }+t\ \left( \chi _{l+1,m}^{\pm
}+\chi _{l-1,m}^{\pm }+\chi _{l,m+1}^{\pm }+\chi _{l,m-1}^{\pm
}\right) \mp  t_{so}\left( \chi _{l+1,m}^{\mp }-\chi _{l-1,m}^{\mp
}\right) -it_{so}\left(
\chi _{l,m+1}^{\mp }-\chi _{l,m-1}^{\mp }\right) =0
\end{equation}

\noindent Here we defined $W_{l,m}^\sigma =\epsilon
+V_{l,m}.$ Now, if we set the on-site energy $\epsilon
=-4t$, and since a uniform (along the y-axis) confinement potential
is assumed, $V_{l,m} \rightarrow V_l$, and thus,
$W_{l,m}^\sigma
\rightarrow W_l$. Hence the equation above can be written in a
matrix form as follows

\begin{equation}
\ ({\bf H}_o-E){\bf A}_m^{\pm }+{\bf T}_{so}^{\pm }{\bf A}_m^{\mp }+
t({\bf A}_{m+1}^{\pm }+{\bf A}_{m-1}^{\pm
})+it_{so}({\bf A}_{m-1}^{\mp }-{\bf A}_{m+1}^{\mp })=0
\end{equation}

\noindent Here we have defined the vector ${\bf A}_m^\sigma$ as the
$mth$ column

\begin{equation}
{\bf A}_m^{\sigma} =\left(
\begin{array}{c}
\chi_{1,m}^{\sigma} \\
\chi_{2,m}^{\sigma} \\
\vdots \\
\chi_{l,m}^{\sigma} \\
\vdots \\
\chi_{N_x,m}^{\sigma}
\end{array}
\right) ,
\end{equation}

\noindent and the ${\bf H}_o$ and ${\bf T}_{so}^{\pm }$ are
$N_x\times N_x$ matrices given by

\begin{equation}
{\bf H}_o=\left(
\begin{array}{cccccc}
W_1 & t & 0 & 0 & \cdots & 0 \\
t & W_2 & t & 0 & \cdots & 0 \\
0 & t & W_3 & t & \cdots & 0 \\
0 & 0 & t & \ddots & \ddots & 0 \\
\vdots & \vdots & \vdots & \ddots & \ddots & t \\
0 & 0 & 0 & 0 & t & W_{N_x}
\end{array}
\right) ,\text{and }{\bf T}_{so}^{\pm }=\left(
\begin{array}{cccccc}
0 & \mp t_{so} & 0 & 0 & \cdots & 0 \\
\pm t_{so} & 0 & \mp t_{so} & 0 & \cdots & 0 \\
0 & \pm t_{so} & 0 & \mp t_{so} & \cdots & 0 \\
0 & 0 & \pm t_{so} & \ddots & \ddots & 0 \\
\vdots & \vdots & \vdots & \ddots & \ddots & \mp t_{so} \\
0 & 0 & 0 & 0 & \pm t_{so} & 0
\end{array}
\right) .
\end{equation}

Now, since both the confinement potential and the spin-orbit
interaction are assumed to be uniform along the longitudinal
direction (propagation direction), the system has  translational
symmetry along the y-axis. Therefore we can make use of the Bloch
theorem. We can define then ${\bf A}_{m+1}^\sigma =e^{ik_ya}{\bf
A}_m^\sigma $ and ${\bf A}_m^\sigma
=e^{ik_ya}{\bf A}_{m-1}^\sigma$, hence ${\bf A}_{m+1}^\sigma=\lambda
{\bf A}_m^\sigma $, with $\lambda =e^{ik_ya}$. Inserting this last
expression for ${\bf A}_{m+1}^\sigma$ in Eq. (A2), we get

\begin{equation}
\ ({\bf H}_o-E){\bf A}_m^{\pm }+{\bf T}_{so}^{\pm }{\bf A}_m^{\mp }+
\lambda t{\bf A}_m^{\pm }+t{\bf A}_{m-1}^{\pm }+it_{so}{\bf
A}_{m-1}^{\mp }-i\lambda t_{so}{\bf A}_m^{\mp }=0,
\end{equation}

\noindent which can be rewritten in a matrix form as follows,

\begin{equation}
\left( \
\begin{array}{cccc}
{\bf H}_o-E & {\bf T}_{so}^{+} & t & it_{so} \\
{\bf T}_{so}^{-} & {\bf H}_o-E & it_{so} & t \\
1 & 0 & 0 & 0 \\
0 & 1 & 0 & 0
\end{array}
\right) \left(
\begin{array}{c}
{\bf A}_m^{+} \\ {\bf A}_m^{-} \\ {\bf A}_{m-1}^{+} \\
{\bf A}_{m-1}^{-}
\end{array}
\right) =\lambda \left(
\begin{array}{cccc}
-t & it_{so} & 0 & 0 \\
it_{so} & -t & 0 & 0 \\
0 & 0 & 1 & 0 \\
0 & 0 & 0 & 1
\end{array}
\right) \left(
\begin{array}{c}
{\bf A}_m^{+} \\ {\bf A}_m^{-} \\ {\bf A}_{m-1}^{+} \\
{\bf A}_{m-1}^{-}
\end{array}
\right) .
\end{equation}

By solving the $(4N_x)\times(4N_x)$ generalized eigenvalue problem,
for a given incident Fermi energy $E,$ all the eigenvalues
$\lambda$ are determined, and hence all $k_y(E),$ yielding the
desired subband spectrum.

\twocolumn

\onecolumn

\begin{figure} 
\epsfysize=6.in 
\epsfxsize=3.in 
\epsfbox{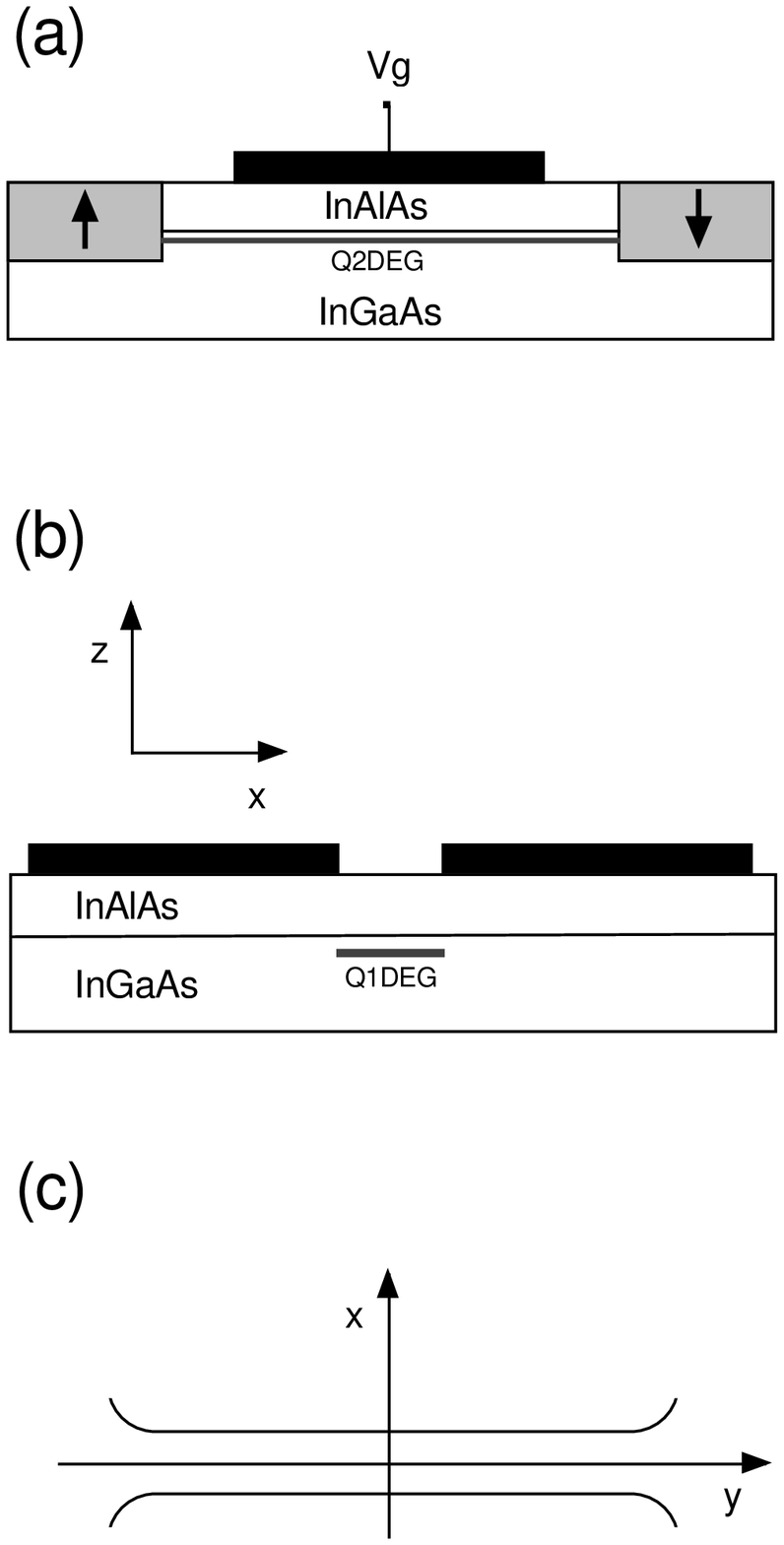}
\caption {(a) Cross sectional schematic of Datta-Das spin modulator
device, (b) cross sectional schematic  of a split gate device used
to create a Q1DEG, (c) diagram of the quasi-one dimensional quantum
channel.}
\end{figure}

\begin{figure} 
\epsfysize=7.in 
\epsfxsize=5.in 
\epsfbox{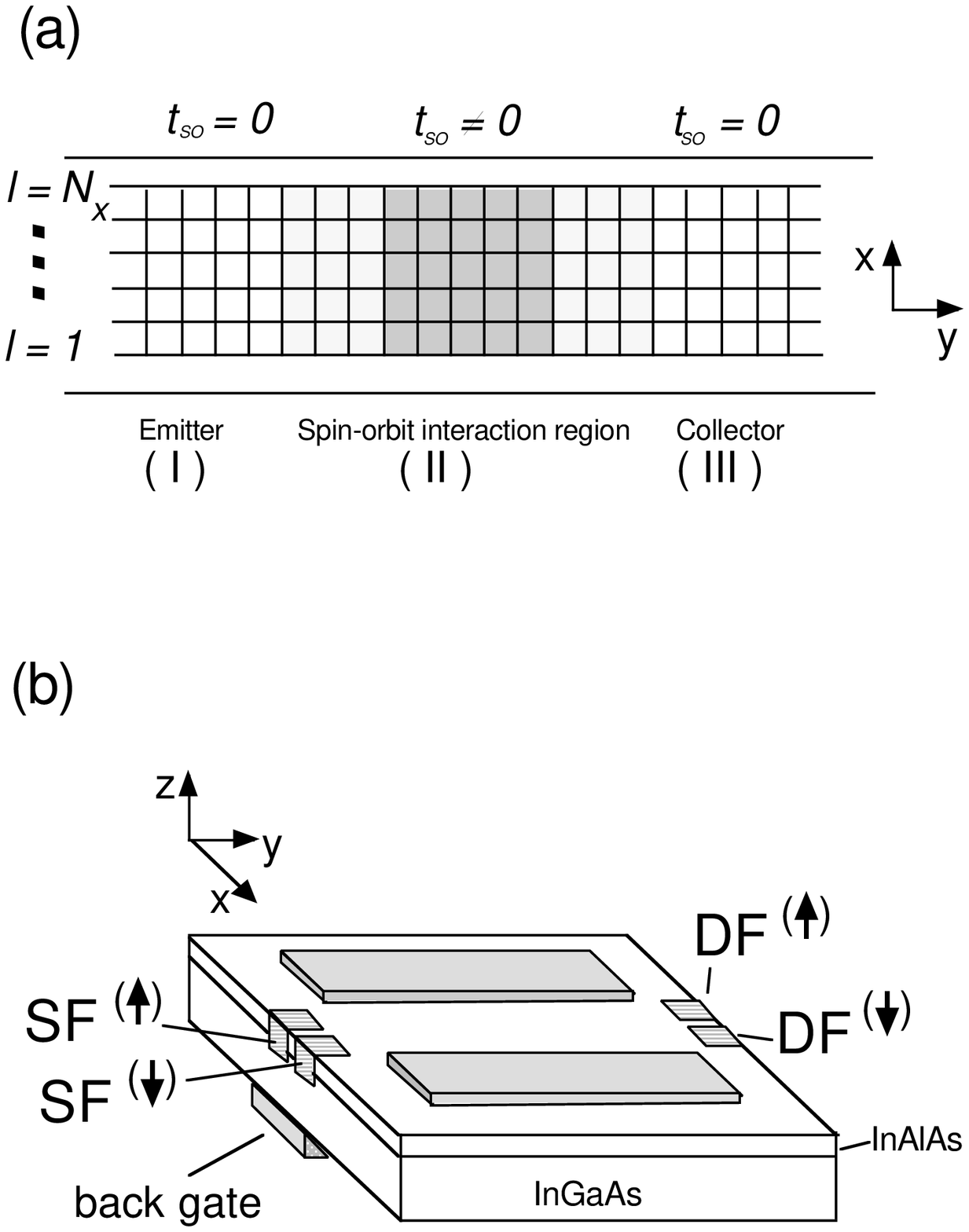}
\vspace{.2in}
\caption{ (a) Schematic of the tight-binding model for the system. In
the shaded areas the spin-orbit interaction is finite, $t_{so}\ne
0$. (b) Schematic of a quasi-one dimensional spin modulator device.
}
\end{figure}

\begin{figure} 
\epsfysize=5.5in 
\epsfxsize=5.5in 
\epsfbox{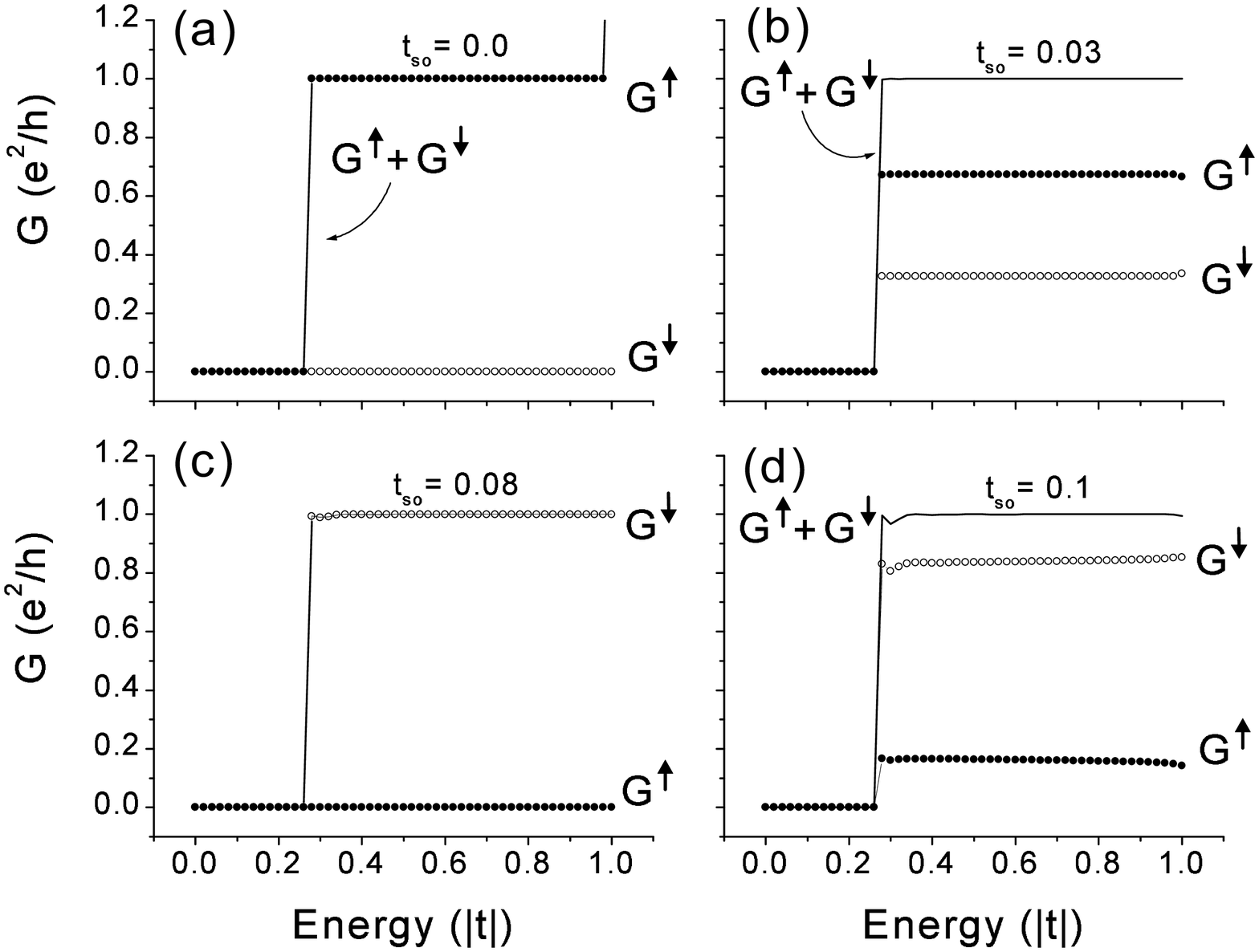}
\caption{Ballistic spin-conductance in units of $e^2/h$ against
the Fermi energy for a narrow quantum wire of width $W=6a$, showing
the spin-modulation for different values of the strength for the
spin-orbit hopping parameter. Filled circles $G^{\uparrow}$, empty
circles $G^{\downarrow}$, and solid line
$G^{\uparrow}+G^{\downarrow}$. In all cases (as in the rest of the
calculations), exclusively spin-up polarized electrons are injected
to the wire. Note that the spin-conductance modulation  is almost
independent of the incident Fermi energy as in the model of Datta
and Das (Ref. [1]).}
\end{figure}

\begin{figure} 
\epsfysize=6.0in \epsfxsize=5.0in \epsfbox{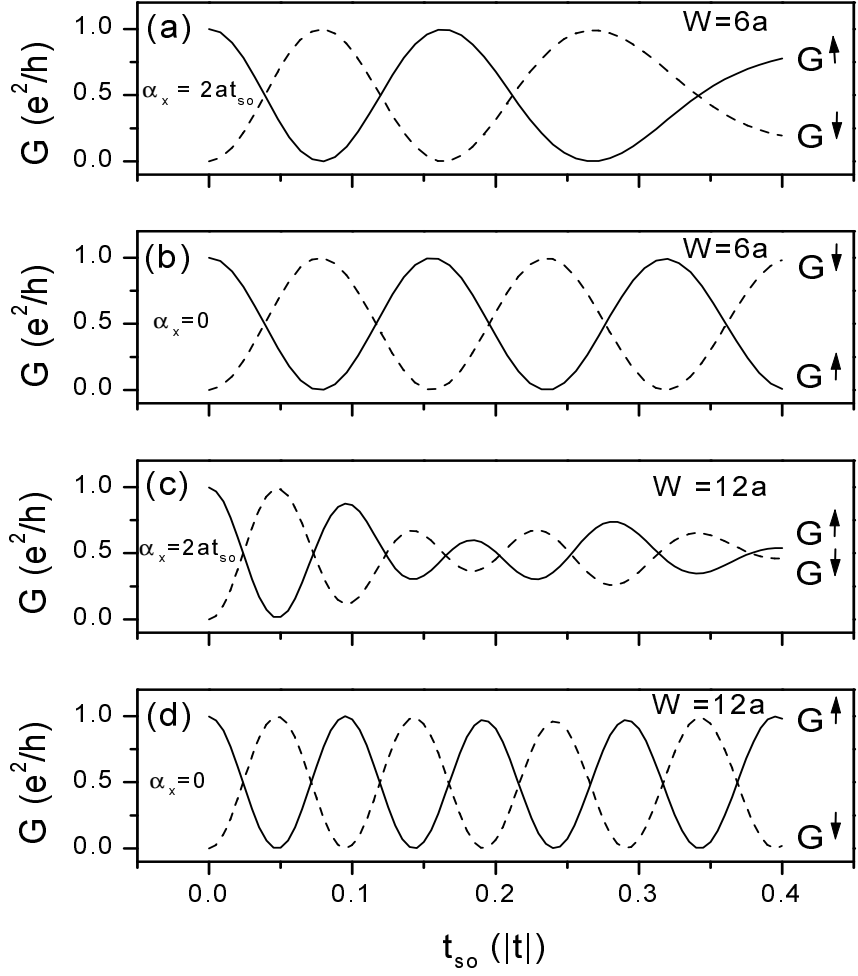}
\caption{Spin-orbit coupling strength dependence of the ballistic
spin-conductance, solid line $G^{\uparrow}$, dashed line
$G^{\downarrow}$: (a) Narrow wire of $W=6a$, and uniform
spin-orbit coupling ($\alpha_x=\alpha_y=2at_{so}$). (b) Same as in
(a) but with $\alpha_x=0$ and $\alpha_y=2at_{so}$, perfect
oscillations are seen for all $t_{so}$. (c) Same as in (a) with
$W=12a$. (d) Modulation for $W=12a$, with $\alpha_x=0$ and
$\alpha_y=2at_{so}$. The intersubband mixing clearly changes the
otherwise perfectly sinusoidal spin-conductance modulation.}
\end{figure}

\begin{figure} 
\epsfysize=4.0in \epsfxsize=5.in \epsfbox{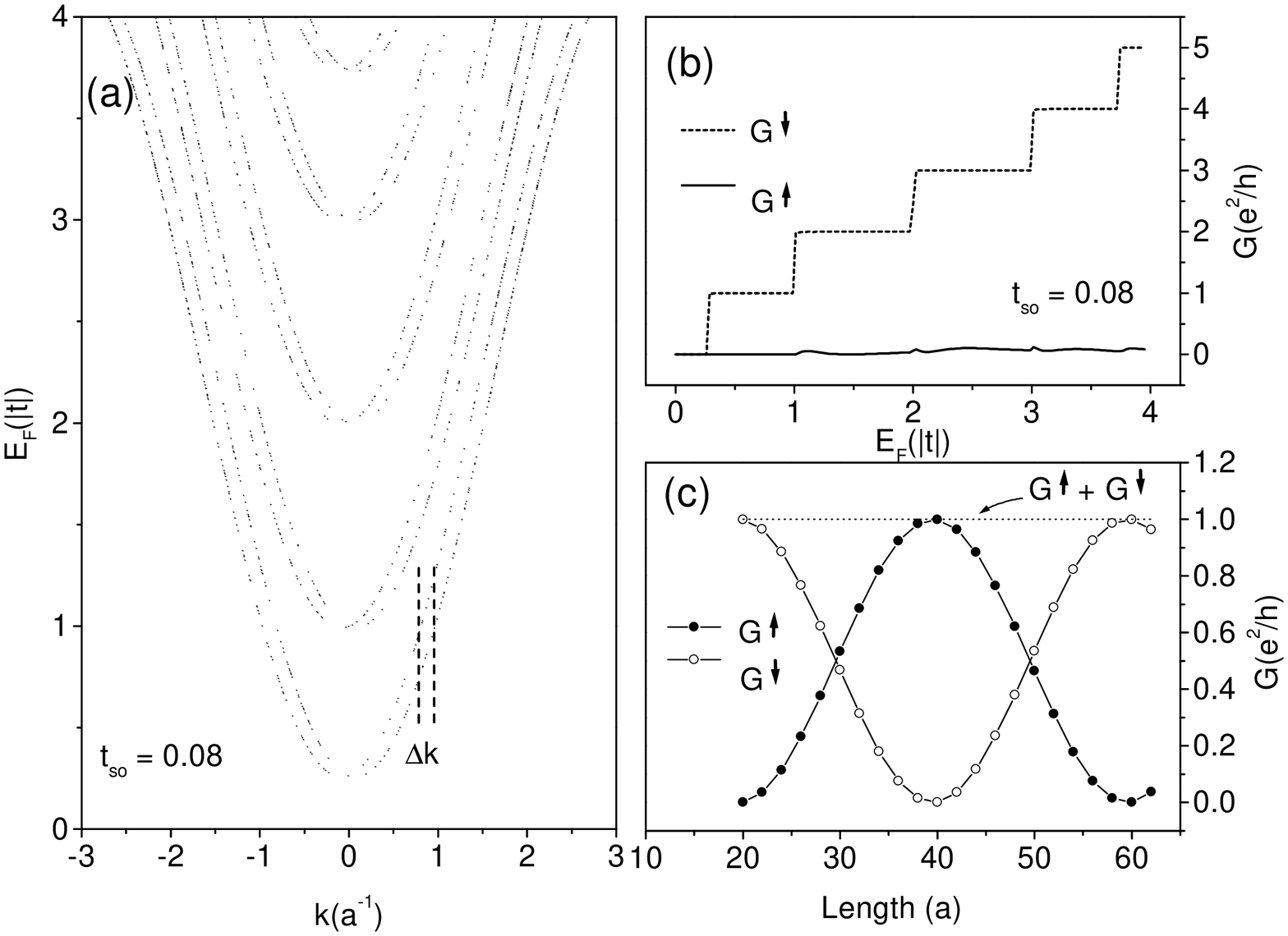}
\caption{(a) Energy subband dispersion for a narrow wire of
$W=6a$ and $t_{so}=0.08$ ({\it weak} spin-orbit coupling regime).
(b) Ballistic spin-conductance for the wire in (a), note that
for this value of $t_{so}$, practically all the conductance is
due to the precessed spin-down electrons (dashed line), while
that for  the spin-up is almost zero for all energies (solid line).
(c) Oscillating behavior of the spin-conductance versus the
effective length of the spin-orbit interaction region, here
$t_{so}=0.08$ as well.  A wave length of $\lambda = 20a$ is
extracted from the data. }
\end{figure}

\vspace{1.55in}
\begin{figure} 
\epsfysize=2.0in \epsfxsize=3.0in \epsfbox{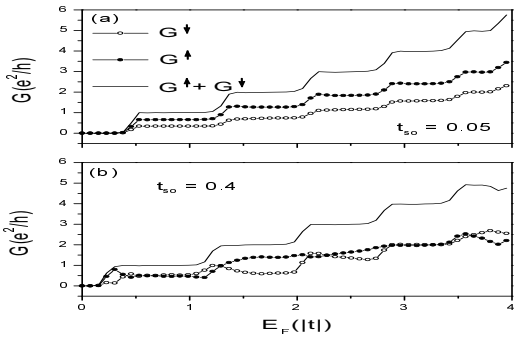}
\caption{ Fermi energy dependence of the ballistic spin-conductance
(in units of $e^2/h$) for a wire of width $W=12a$ and confinement
strength (curvature) of the parabolic potential $w_x=0.2$.
(a) {\it Weak} spin-orbit coupling case, $t_{so}=0.05$. (b) {\it
Strong} spin-orbit coupling case, $t_{so}=0.4$. Filled circles
$G^{\uparrow}$, empty circles $G^{\downarrow}$, and solid line
$G^{\uparrow}+G^{\downarrow}$.  }
\end{figure}

\begin{figure} 
\epsfysize=3.0in \epsfxsize=4.in \epsfbox{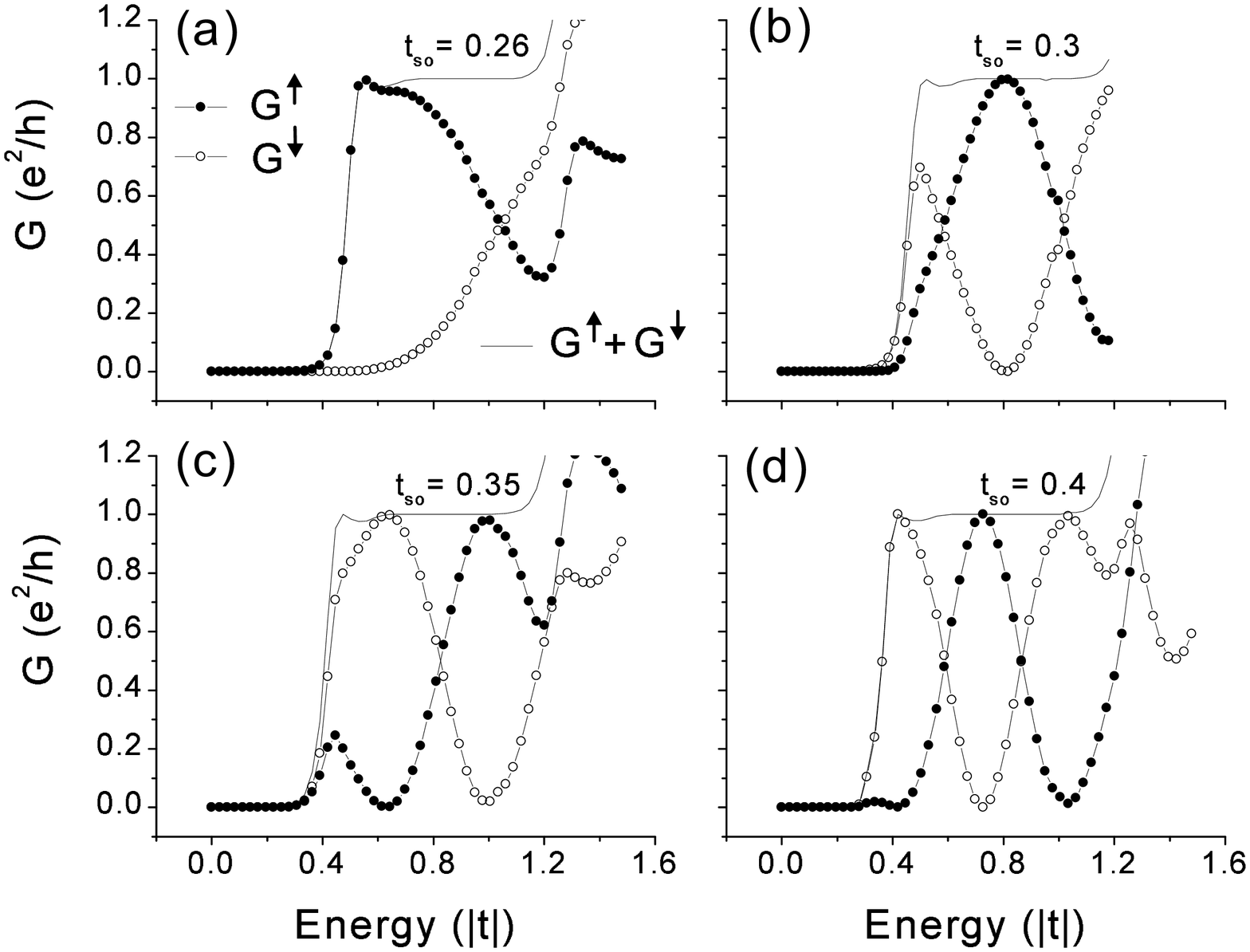}
\caption{Ballistic spin-conductance for the {\it strong} spin-orbit
coupling regime, ($t_{so} > 0.22 $) for the wire of Fig. 3. Only the
first propagating mode for each spin is shown. In this regime, both
$G^{\uparrow}$ and $G^{\downarrow}$ are strongly dependent on the
incident Fermi energy, with an oscillating behavior as $t_{so}$ is
enhanced, which exhibits the energy dependence of the
spin-precession.  Cases (c) and (d) are particularly interesting.
For $E_F\approx 1.0$ in (d) the polarity of the transmitted
electrons is reversed (with respect to the injected spin-up
polarized electrons) Perfectly polarized spin up or spin down
electrons can be emitted from the wire depending on the Fermi
energy. }
\end{figure}

\begin{figure} 
\epsfysize=4.0in \epsfxsize=5.in \epsfbox{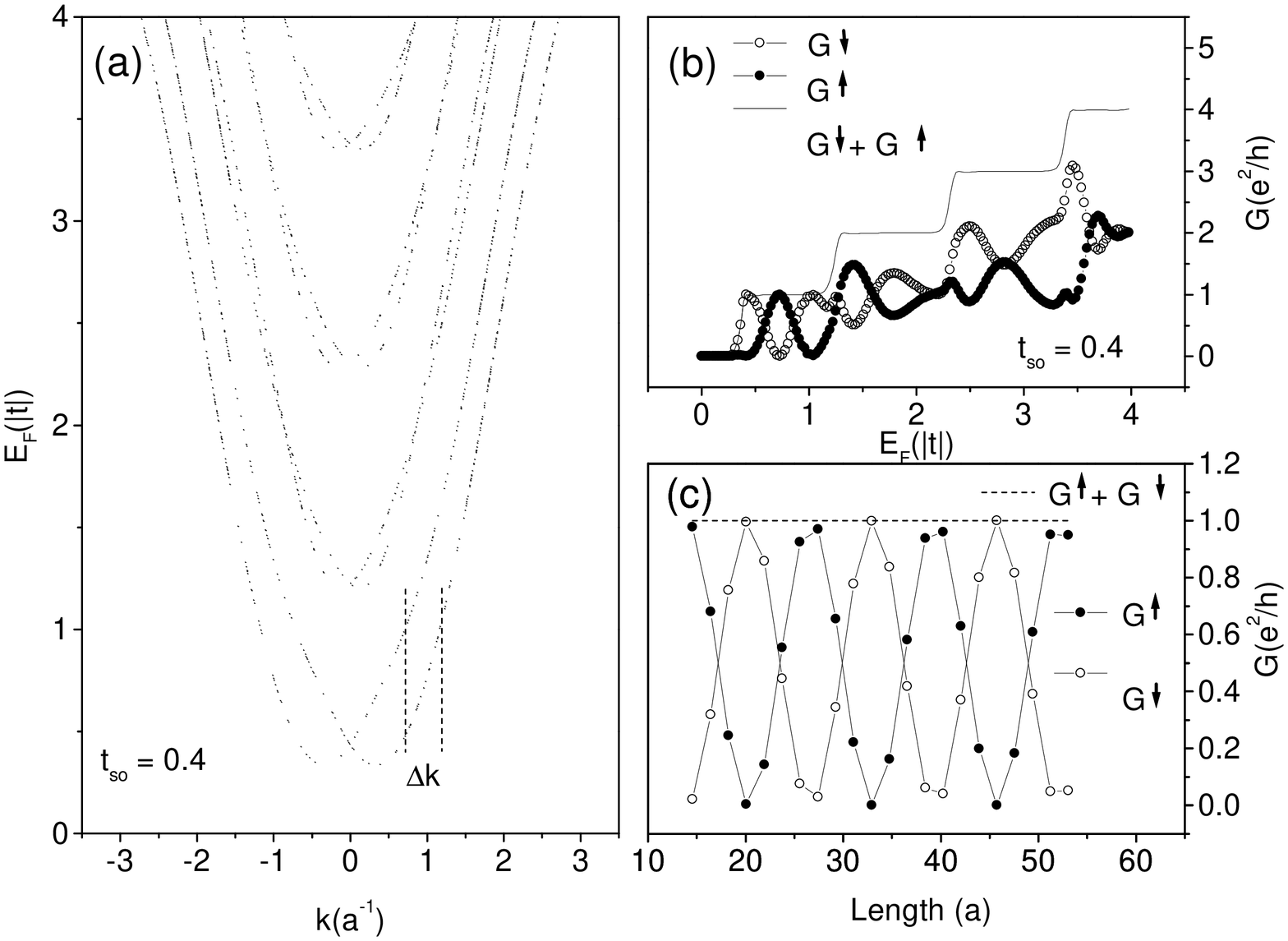}
\caption{(a) Energy subband dispersion for a narrow wire of
$W=6a$ and $t_{so}=0.4$ ({\it strong} spin-orbit coupling regime).
Note that the subbands are no longer parabolic as in {\it weak}
$t_{so}$ regime, resulting in a $\Delta k \rightarrow \Delta
k(E)$. (b) Ballistic spin-conductance for the wire in (a);
filled  circles $G^{\uparrow}$, empty circles $G^{\downarrow}$,
and solid line
$G^{\uparrow}+G^{\downarrow}$. The energy dependence is clearly
evident here. (c) Oscillating behavior of the spin-conductance versus
the effective length of the spin-orbit interaction
region ($t_{so}=0.4$) . The wave length of the oscillations is
$\lambda \approx 13.8a$. }
\end{figure}

\begin{figure} 
\epsfysize=2.5in \epsfxsize=2.5in \epsfbox{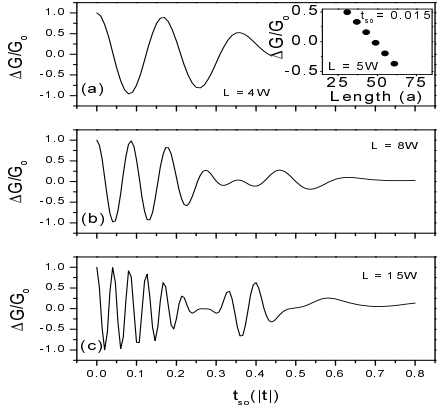}
\caption{Plots of the relative conductance change
[$\Delta G/G_o=(G^{\uparrow}-G^{\downarrow})/
(G^{\uparrow}+G^{\downarrow})$] against the strength of spin-orbit
parameter for three different effective lengths of the spin-orbit
(Rashba) interaction, $L=4W$, $L=8W$, and $L=15W$, with $W=6a$.
Clearly the beat-like behavior is enhanced with $L$, showing the
nature of the electron spin-precession induced by the Rashba
effect. In the inset of (a) the relative conductance change is
plotted as a function of the effective length for a {\it weak}
spin-orbit hopping parameter, $t_{so}=0.015$. A change in sign for
$\Delta G/G_o$ as the length is increased is evident showing that
the effect of the spin-precession still important for such regime.
}
\end{figure}


\begin{references}
\bibitem{tso-values} For the currently typical experimental values of
the spin-orbit Rashba parameter, the spin-orbit hopping parameter is
estimated to be in the range between $0.001|t|$ to $0.015|t|$,
approximately.
\bibitem{phase}  The differential phase shift Eq. (2), is rewritten
in terms of the strength of the spin orbit coupling by $\Delta \theta
= 2\beta_{so}\left (\frac{L}{a}\right)$ where
$\beta_{so}=t_{so}/|t|$ is the strength of spin-orbit hopping
parameter in units of $|t|$.
\bibitem{SdH-like} In contrast to Shubnikov-de Haas (SdH)
oscillations the beating in Fig. 9 is a quantum interference
effect. If a large number of channels is present more
complicated patterns emerge.
\end{references}
\end{document}